\documentstyle[psfig]{l-aa}
\input psfig.sty

\newcommand{\Cap}{{\rm Ca\,II}}
\newcommand{\Tip}{{\rm Ti\,II}}
\newcommand{\Fep}{{\rm Fe\,II}}
\newcommand{\Mnp}{{\rm Mn\,II}}

\newcommand{\Capp}{{\rm Ca\,III}}
\newcommand{\Mgp}{{\rm Mg\,II}}
\newcommand{\Mgo}{{\rm Mg\,I}}
\newcommand{\Ho}{{\rm HI}}
\newcommand{\Hy}{{\rm H}}
\newcommand{\X}{{\rm X}}
\newcommand{\Fe}{{\rm Fe}}
\newcommand{\Mn}{{\rm Mn}}
\newcommand{\Ti}{{\rm Ti}}

\newcommand{\h}{{\rm H}}
\newcommand{\dla}{{\rm dla}}
\newcommand{\obs}{{\rm obs}}
\newcommand{\ism}{{\rm ism}}

\def\mincir{\ \raise -2.truept\hbox{\rlap{\hbox{$\sim$}}\raise5.truept  
\hbox{$<$}\ }}                                                          
\def\magcir{\ \raise -2.truept\hbox{\rlap{\hbox{$\sim$}}\raise5.truept  
\hbox{$>$}\ }}                                                          
 
\def\asec{$^{\prime\prime}$ }           

\begin{document}
\thesaurus{11.17.1 - 11.09.4 - 11.01.1 - 11.02.2 - 10.01.1 }

\title{
The $z=0.558$ absorption system towards PKS 0118-272: ~~~~~~
A candidate Damped Ly $\alpha$ system at low redshift
\thanks{Based on observations collected at the European Southern Observatory 
(La Silla, Chile)}
}

\author{G. Vladilo  \inst{1},  
M. Centuri\'on \inst{2}, R. Falomo \inst{3}, P. Molaro \inst{1} 
 }

\offprints{ G. Vladilo,
Osservatorio Astronomico di Trieste, Via G.B. Tiepolo 11,
        34131,  Trieste, Italy
}

\institute{        
        Osservatorio Astronomico di Trieste, Via G.B. Tiepolo 11,
        34131, Trieste, Italy;  
        e-mail: vladilo@ts.astro.it, molaro@ts.astro.it
\and 
        Instituto de Astrof\'\i sica de Canarias, 38200, La Laguna
        (Tenerife), Spain;
        e-mail: mcm@iac.es
\and
	Osservatorio Astronomico di Padova, Vicolo dell' Osservatorio 5,
	35122 - Padova, Italy;
	e-mail: rfalomo@pd.astro.it
}

\date{Received date; accepted date}

\maketitle

\markboth
{Vladilo et al.: Candidate damped Ly\,$\alpha$ system at $z=0.558$ towards PKS 0118-272}
{Vladilo et al.: Candidate damped Ly\,$\alpha$ system at $z=0.558$ towards PKS 0118-272}

\begin{abstract}

We present a study of the $\Mgp$ absorption system 
at $z$ = 0.558 towards the BL Lac object PKS 0118--272
based on
high resolution spectra
($\lambda/\Delta\lambda  \simeq 2 \times 10^4$)
obtained  at the 3.6m ESO  telescope
and  on direct imaging data obtained  
at the 3.5m ESO New Technology Telescope. 
At the redshift of the absorber we detect   
lines of the  low ionization  species  $\Mgo$, $\Mgp$, $\Cap$, $\Tip$,
$\Mnp$, and $\Fep$. 
Most of the lines are unsaturated and we determine
accurate column densities for all the species but $\Mgp$. 
The derived column densities  are typical of  the  interstellar gas
in the disk of the Galaxy. 
By assuming dust-free gas with solar abundances 
these column densities yield
$N$($\Ho$) $>$ 2.5 $\times$ 10$^{19}$ cm$^{-2}$;
however, the  relative elemental abundances
suggest  that some dust is present and that   
$N$($\Ho$) $\simeq$  2  $\times$ 10$^{20}$ cm$^{-2}$.
The inferred $\Ho$ column density indicates that the absorber 
is a Damped Ly\,$\alpha$ system.
We find [Ti/Fe] = +0.3, in agreement with [Ti/Fe]  measurements
in DLA absorbers, but [Mn/Fe] = +0.4, at variance with the
values [Mn/Fe] $<$ 0 common to DLA systems.  
The measured [Ti/Fe] and [Mn/Fe] ratios match remarkably well 
the differential depletion pattern   of
low-density interstellar clouds in the Galaxy. 
Unlike high-redshift DLA systems ($z \geq 2$), 
the $z$=0.558 absorber seems to originate in a galaxy  
that has already attained the abundances and dust content
of present-day disk galaxies.  
The analysis of our imaging data lends support
to the presence of an intervening galaxy. 
After a careful subtraction  of the BL Lac image, 
an object at 1.6\asec\ from PKS 0118--272 is detected. 
At the absorber redshift the
projected distance of this close companion (14 $h_{50}^{-1}$ kpc) 
and its absolute magnitude 
(M$_R \approx$ --22.3)  are consistent with those found  
for galaxies associated with low-redshift DLA systems.

\keywords{Quasars: absorption lines - Galaxies: ISM, abundances -
BL Lacertae objects: individual: PKS 0118--272 - Galaxy: abundances}
\end{abstract}

\section{Introduction}

Damped Lyman $\alpha$ (DLA) systems are the class of QSO absorbers
with highest neutral hydrogen column density, 
$N$($\Ho$) $\geq$ 2 $\times$ 10$^{20}$ cm$^{-2}$, comparable
with that observed in  present-day disk  galaxies
(Wolfe et al. 1986). 
The comoving mass density of neutral gas in DLA systems,
$\Omega_g(z)$,  is a significant fraction of the
barionic content of the high redshift universe; 
$\Omega_g(z)$  shows a decrease with time which 
is interpreted as  gas consumption by star formation
(Wolfe et al. 1995). 
The degree of ionization and the velocity dispersion of the
metal lines in the DLA spectra  
resemble the quiescent neutral gas residing in rotating galactic disks. 
For all the above reasons DLA systems   most likely originate 
in galactic (or proto-galactic) $\Ho$  regions  and their study 
can be used to trace the evolution of galaxies ---
via their gas component --- up to the highest redshift 
QSOs. 

About 90 DLA systems have been currently identified,
covering the redshift range  $0.2 \leq z_{\rm abs} \leq 4.4$
(Wolfe et al. 1995, Storrie-Lombardi et al. 1996).  
Owing to the paucity
of QSO spectroscopic studies in the ultraviolet, however, only 6 
damped Ly $\alpha$ absorptions at redshift $z_{\rm abs} \leq 1.7$
have been observed, either with the {\it IUE} (Lanzetta et al. 1995)
 or with the {\it HST}
(Cohen et al. 1994, Steidel et al. 1995). 
No  DLA systems at  $z_{\rm abs} \leq 1.3$ have been as yet found in the
{\it HST} QSO absorption-line key project survey (Bahcall et al. 1996) and 
only a few DLA systems at moderate redshift are known from 21 cm 
$\Ho$ observations (see Table 4 in Wolfe et al. 1995 and references
quoted therein). 

The small fraction of low-redshift DLA systems  
limits the possibility of tracing their evolution  
up to a look-back time of about two thirds of the age of the universe.
It is therefore of interest to find new DLA systems at moderate
redshift in order to fill this large temporal gap. 
Moreover, only at low redshift is it possible to perform
a  morphological  identification of the intervening galaxies
via direct imaging techniques. 
A study of  
galaxies associated with DLA absorbers at $z \leq 1$ 
in the sightlines of seven quasars has been 
performed  by  Le Brun et al.\ (1997), who find
a wide variety of morphologies and
a large spread of luminosities.  
 
High redshift DLAs  are now being systematically investigated 
at high spectral resolution
(see Lu et al. 1996,   Molaro et al. 1997,
Pettini et al. 1997, Prochaska \& Wolfe 1997  and refs. quoted
in these papers). 
However, only a very few DLA systems at moderate redshift 
have been observed at $\lambda/\delta\lambda \geq 10^4$
(Lanzetta \& Bowen 1992; Meyer et al. 1995; Prochaska \& Wolfe 1997). 
 Here we show that the absorption system
at $z_{\rm abs}$ = 0.558 towards PKS 0118-27,  
originally detected in $\Mgp$ by Falomo (1991)
at  $\lambda/\delta\lambda \simeq  10^3$
is most likely a DLA system. 
We present new spectroscopic ($\lambda/\delta\lambda \simeq 2 \times 10^4$)
and imaging observations
of this system  in Sect. 2. 
The absorption spectrum is analysed in Sect. 3,
while in Sect. 4 we discuss the spectroscopic evidence that the absorber
is a DLA system.  The positive results of a search for an intervening
galaxy are reported in Sect. 5. 
 The properties of the absorber are compared
with those of DLA systems and of Galactic
interstellar clouds in Sect. 6. 
The results are summarized in Sect. 7.

\section{Observations and data reduction}

The optical spectra of PKS 0118$-$272  ($m_{\rm R}$ = 15.9 mag) 
 were obtained 
with the  CASPEC spectrograph (Pasquini \& D'Odorico 1989)  
at  the ESO 3.6m telescope (La Silla, Chile).  
The 31.6 lines/mm echelle grating   and the
long camera (f/3) were used.  
The detector was a thin, back-illuminated Tektronics TK1024AB CCD
with 1024x1024 square pixels 24 $\mu$m in size. 
The CCD was binned at a step of 2 pixels along the dispersion, 
yielding a binned pixel size of 0.134 \AA\
at $\lambda$ = 500 nm. The slit width was fixed at 300 $\mu$m
(2.1$\arcsec$) and the resulting resolving power was $R 
= \lambda / \Delta\lambda_{\rm instr} \simeq 19500$,
which represents a factor of 20 increase in resolution with respect to
previous observations of the same object. 
The full width at half
maximum of the instrumental profile, $\Delta\lambda_{\rm instr}$
was measured from
  the emission lines of the Thorium-Argon arcs recorded for 
each spectrum. 
A log of the observations including the central wavelengths
of the different spectral ranges covered is given in Table 1. 
The spectra were reduced using the echelle reduction
package implemented within MIDAS. 
The uncertainty in the wavelength calibation is 
$\delta\lambda$ $\simeq$ 0.05 \AA, corresponding to 0.2
of a resolution element. 
The signal-to-noise ratio per pixel, defined as the inverse relative
${\it rms}$ scatter of the continuum, is typically
S/N $\simeq$ 20. The 3-$\sigma$ threshold equivalent width for detection
of narrow absorption lines is $W_{\rm lim}$ $\simeq$ 40 m\AA,
corresponding to $\simeq$ 25 m\AA\ in the rest frame.  
The spectral ranges of the final, normalized spectrum
including the identified lines of the $z=0.558$ absorber 
are shown in Fig. 1.

\begin{table}
\caption{Observations of PKS 0118$-$272.}
\begin{flushleft}
\begin{tabular}{cccc}
\hline 
 Date     & $\lambda_c$ & Exp. time & No. of    \\
          & (nm)        &  (s)      & spectra   \\
\hline
14-09-95 &   495       & 3000      &   2        \\
15-09-95 &   395       & 3300      &   2        \\
16-09-95 &   570       & 3600      &   2        \\
\hline 
\end{tabular}
\end{flushleft}
\end{table}

Optical images were obtained with the 3.5m New Technology Telescope
(NTT) at  ESO using the direct
imaging system SUSI (Melnick et al.\,1992).  Data were acquired using an
R-band filter (Cousins system) and a CCD (TK 1024) with 24$\mu$m pixel
size corresponding to 0.13\asec on the sky.  Conditions were
photometric and seeing was 0.9\asec (FWHM).  Observations of standard
stars  were used to set the photometric zero point.
The images  were processed in the standard way  using  
IRAF procedures. 
Images of PKS 0118-272 were taken in August 1992 and March
1993. In the second case several short (2 min) exposure frames
were obtained.  These images were then co-added to obtain an
average frame.  The two data sets have similar resolutions and yield 
practically identical results.
The central part of the final average image  is shown in Fig. 2.

\begin{table}
\begin{flushleft}
\caption{Absorption lines of the $z_{\rm abs}$=0.558 system.}
\begin{tabular}{llllc}
\hline
Ion &  $\lambda_{\rm lab}$ & $\lambda_{\rm obs}$ & W$_{\lambda}$ & $\sigma$ \\
    &      (\AA)           &   (\AA)             & (\AA)     &  (\AA)  \\
\hline 
$\Cap$  & 3934.78   & 6130.40 & 0.103 &0.011 \\
$\Cap$  & 3969.59   & 6184.64 & 0.058& 0.006 \\
$\Tip$ & 3384.74   & 5273.44 &  0.045 & 0.008 \\
$\Fep$ & 2344.21   & 3652.27 &  0.270& 0.021\\ 
$\Fep$ & 2374.46   & 3699.39& 0.160& 0.020    \\
$\Fep$ & 2382.76   & 3712.33& 0.340& 0.020 \\
$\Fep$ & 2586.65   & 4029.98& 0.263& 0.015\\
$\Fep$ & 2600.17   & 4051.05& 0.359& 0.014\\
$\Mnp$ & 2576.88   & 4014.76 & 0.071& 0.015\\
$\Mnp$ & 2594.50   & 4042.21 &  0.058& 0.014 \\
$\Mnp$ & 2606.46   & 4060.85 &  0.039& 0.014 \\
$\Mgo$ & 2852.96   & 4444.92 & 0.160& 0.012 \\
$\Mgp$ & 2796.35   & 4356.69 & 0.494 & 0.011 \\
$\Mgp$ & 2802.53   & 4367.89 & 0.462& 0.011\\
\hline
\end{tabular}
\bigskip
\end{flushleft}
\end{table}

\begin{figure*} 
\psfig{figure=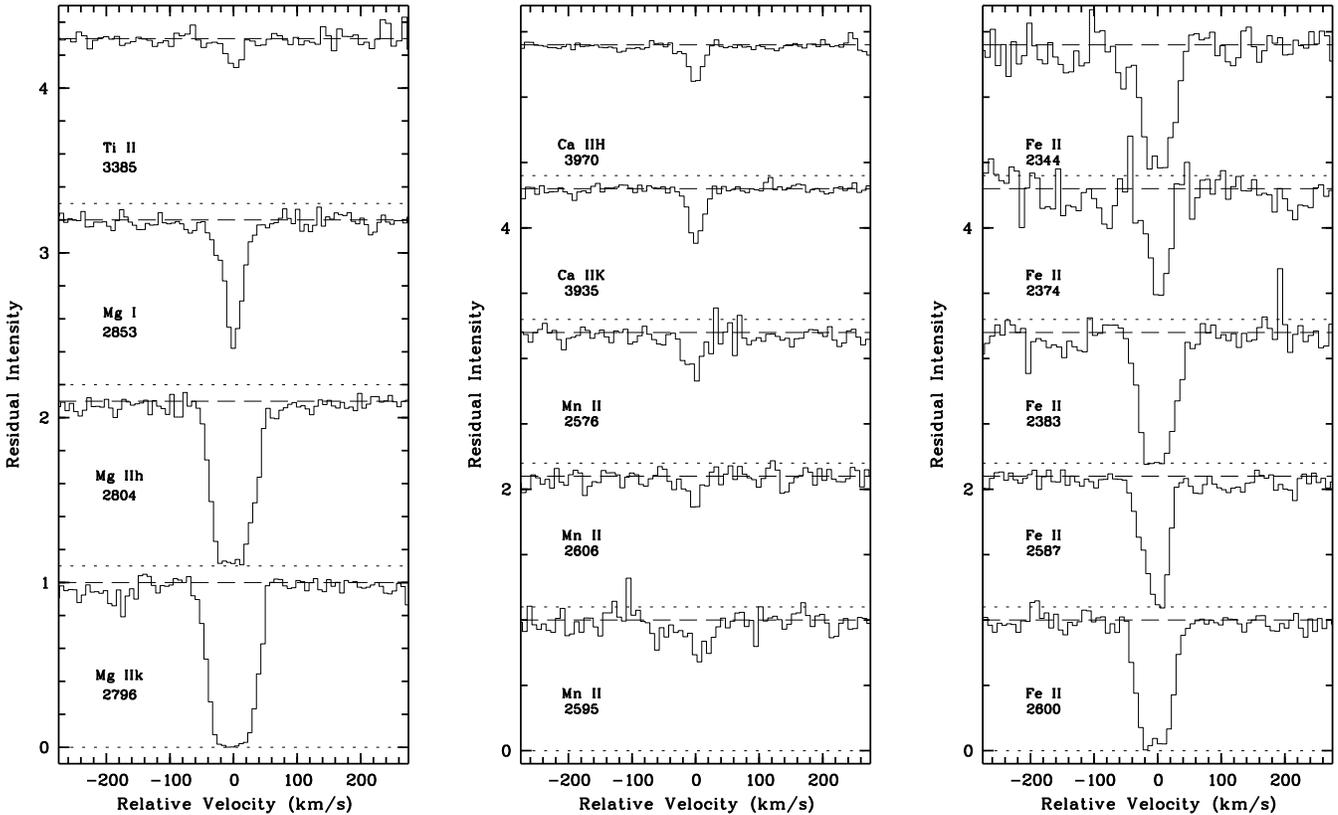,width=17.6cm,angle=-90}
\caption
{Portions of the normalized spectrum of
PKS 0118$-$272. Rest wavelengths of the transitions are indicated in Angstroms.
Velocities are relative to $z_{\rm abs}$ = 0.5580.} 
\end{figure*}

\begin{figure}   
\psfig{figure=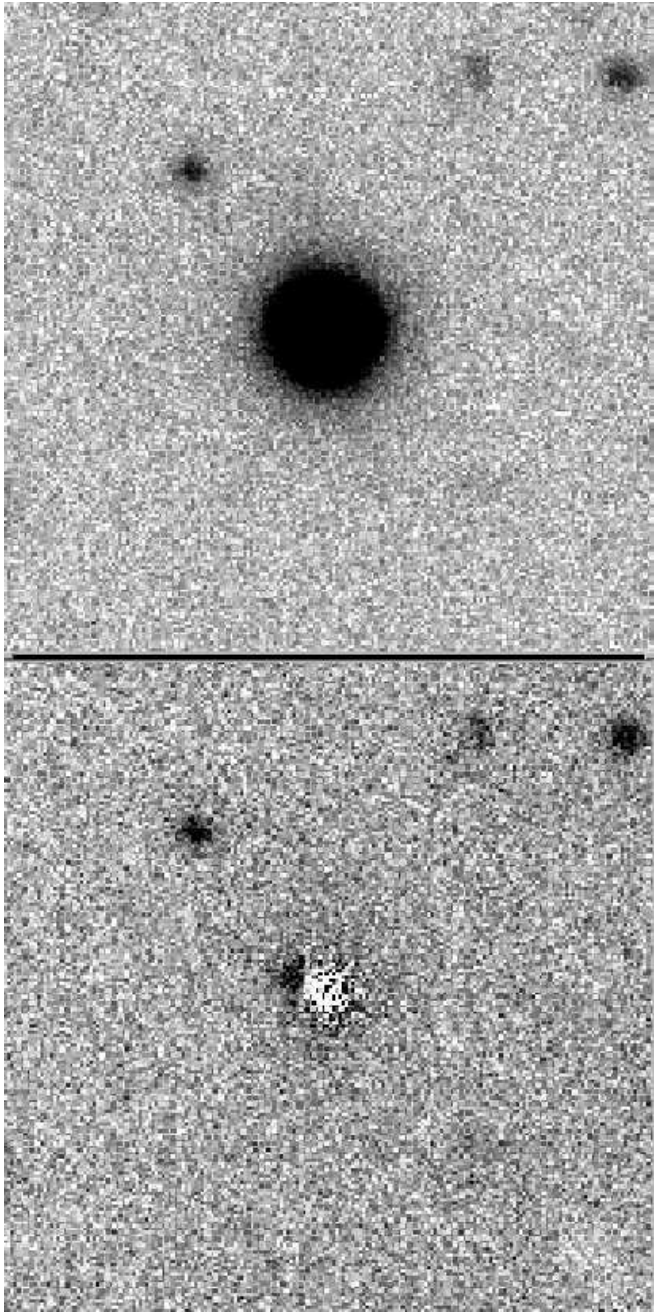,width=8.6cm}
\caption{
{\bf (a)} Upper image: field around  PKS 0118-272 
showing the galaxy   at 8\asec\ (PA $\simeq$ 40$^{\circ}$) from
the  BL Lac object and other galaxies in the field. 
{\bf (b)} Lower image: same field after subtraction
of a scaled PSF template.
A close companion galaxy, similar to the other galaxies in the field,  
 is apparent   1.6\asec\ NE of 0118-272.}
\end{figure}

\section{The absorption spectrum}

A total of 14 absorption lines of  $\Mgo$, $\Mgp$, $\Cap$, $\Tip$,
$\Mnp$, and $\Fep$ 
detected at the redshift of the absorber
are shown in Fig. 1  and listed in  Table 2.
This represents  
the most extensive collection of transitions for
a low redshift  absorption system.
$\Tip$ had been detected previously   only in two QSO absorbers
(Meyer et al. 1995; Prochaska \& Wolfe 1997)
and $\Cap$ detections are also relatively rare
(Robertson et al. 1988; Bowen 1991). 
The  neutral and low-ionization species detected are likely
to be associated with a  neutral hydrogen component 
present in the system.

The observed profiles do not show  multiple component structure
at our resolution (FWHM$\simeq$15 km s$^{-1}$). 
Only a slight asymmetry, barely visible in some lines
(e.g. $\Mgo$ $\lambda$2853\AA\ and $\Fep$ $\lambda$2587\AA),   
suggests the presence of an
additional, weak component in the blue wing of the main one. 
The total  radial velocity extent
of the systems never exceeds 120 km s$^{-1}$,
the value measured in the saturated $\Mgp$ k line.
The kinematics of the system appears to be    simple, 
at least along  our   line of sight. 

In Table 2 we list the rest frame   
equivalent widths and 1 $\sigma$ errors
of the detected lines   of the $z=0.558$ system. 
Equivalent width errors  never exceed 20 m\AA\ in the rest frame.
A significant fraction of the lines, including
$\Cap$ H and K, $\Tip$ $\lambda$3385\AA, and the 3 $\Mnp$ lines,
are almost unsaturated, with $W_{\lambda}$
  $\le 100$ m\AA.
The lines $\Mgo$ $\lambda$2852\AA\ and $\Fep$ $\lambda$2374\AA\ are only mildly
saturated, with  $W_{\lambda}$ $\simeq 160$ m\AA.
The low degree of saturation
guarantees a reliable determination of the column densities,
with the only exception of $\Mgp$, which shows both h and k lines
heavily saturated. 
 
Column densities and $b$ values
were derived by fitting
Voigt profiles convolved with the instrumental point spread function.
The set of routines FITLYMAN (Fontana \& Ballester 1995) included
in the MIDAS package were used. 
Atomic parameters were taken from Morton (1991). 
The results  are shown in Table 3.
Quoted errors are 1 $\sigma$ and result from the
fitting procedure.  
Broadening parameters were derived independently for
the species having at least two transitions ($\Cap$, $\Mnp$
and $\Fep$).
These $b$ values span the range from 16 to 20 km s$^{-1}$, i.e.
much higher than the   values
expected for pure thermal broadening in low-ionization gas,
where $T \leq 10^{4}$ K and $b$  $\leq$ 3 km s$^{-1}$
for any of the metals considered here. This suggests that  turbulence
and/or macroscopic motions
dominate the   broadening. Hence a common $b$ value is expected for ions
tracing   same volumes of gas, independently of their atomic masses.
The $b$ value derived from $\Mnp$ and $\Fep$ are  equal within
the errors, consistently with this interpretation.
The lower $b$ value derived from the $\Cap$ lines is not
in contradiction since $\Cap$ does not trace exactly the same
volumes as $\Mnp$ and $\Fep$     owing to 
differences in ionization potential.
The mean $b$ value of the $\Mnp$ and $\Fep$ was
adopted as representative of the ionized species  $\Tip$ and $\Mgp$. 
The $\Tip$ column density derived in this way is essentially equal
to the one derived by leaving the $b$ parameter free during the
fitting procedure (see Table 3).

\begin{table}
\caption{Column densities of the $z$=0.5580 system.}  
\begin{flushleft}
\begin{tabular}{llll}
\hline
Ion  &  $N$        &   $b$         & comments    \\
     & (cm$^{-2}$) & (km s$^{-1}$) &             \\
\hline 
$\Mgo$ & 12.47 $\pm$ 0.06     &  16.2 $\pm$ 2.1  & best fit 1 line \\
$\Mgp$ & 14.00  $^{+2.30}_{-0.10}$  &  19.2 $\pm$ 1.1 & b from $\Fep$ and $\Mnp$ \\
$\Cap$& 12.37  $\pm$ 0.07   &  11.6 $\pm$  3.4        &  best fit 2 lines \\
$\Tip$ & 12.27  $\pm$ 0.19    &  13.3 $\pm$  9.2        &  best fit 1 line\\
      & 12.31  $\pm$ 0.01    &  19.2 $\pm$ 1.1 & b from $\Fep$ and $\Mnp$ \\
$\Mnp$ & 12.77  $\pm$ 0.06    &  20.1 $\pm$ 3.9  & best fit 3 lines \\
$\Fep$ & 14.45  $\pm$ 0.03    &  18.3 $\pm$ 0.5 & best fit 5 lines\\
\hline
\end{tabular}
\end{flushleft}
\end{table}

\section{Spectroscopic evidence that the $z$=0.558 absorber is a 
Damped Ly\,$\alpha$ system}

The available ultraviolet spectra of PKS 0118--272, taken with the {\it IUE}
satellite, do not cover the Ly $\alpha$ line at $z$=0.558.
This prevents the direct determination of the hydrogen column density,
which would allow to identify the absorber as a DLA system if 
$N$($\Ho$) $\approx$ 10$^{20}$ cm$^{-2}$. 
In order to estimate $N$($\Ho$) we   used
the column densities of the dominant ions of Table 3.
Calcium was omitted  because $\Capp$ can also exist  
in $\Ho$ regions. Magnesium was
excluded owing to its large  column density error.
By assuming solar abundances
and no dust depletion 
we   estimated the lower limits,  $N$($\Ho$)$_{\rm min}$,   shown 
in the second  column
of Table 4. Here and in the rest of this work
solar abundances are taken from Anders \& Grevesse (1989),
with the exception of the abundance of Fe
which was taken from Hannaford et al.
(1992). 
The  conservative limits $N$($\Ho$)$_{\rm min}$   indicate that the  
total column density  is   in excess of    
2.5 $\times$ 10$^{19}$ atoms cm$^{-2}$, the most stringent   limit
coming from $\Mnp$. 

A prediction of the $\Ho$ column density in the presence of dust
can be obtained 
by adopting for each element X a value of interstellar depletion 
$
\delta_{\X,\ism} = \log\,N({\rm X})/N({\rm H}) - \log\,({\rm X/H})_{\sun} ,
$
from Galactic interstellar surveys.  
Here and in the rest of this paper we adopt the representative values 
of depletions from the compilation of Jenkins (1987).  
It is known that the quantities $\delta_{\X,\ism}$ correlate with the
mean density along the line of sight, 
$\overline{n}$(H). Following the
definitions of Jenkins (1987) we consider two situations:
{\it base} depletion, representative of low-density
lines of sight (log\,$\overline{n}$(H) = --1.5)
and {\it dense} depletion, representative
of lines of sight with high mean density (log\,$\overline{n}$(H) = +0.5). 
For each element in Table 4 we give the typical value
of $\delta_{\X,\ism}$ in each case and  the estimated
hydrogen column density  derived by taking
into account the presence of dust,  $N$($\Ho$)$_{\rm est}$.
All the $N$($\Ho$)$_{\rm est}$ values
are well in excess of 10$^{20}$ atoms cm$^{-2}$. 
However, very different and mutually inconsistent estimates 
result from assuming a dense depletion pattern. 
Instead, a remarkable agreement at $\simeq$
2 $\times$ 10$^{20}$ atoms cm$^{-2}$
is found from the 3 independent predictions if
the depletion pattern is typical  
of moderately depleted interstellar clouds. 
In any case, even assuming the above quoted   lower limit,
the high total column density provides indirect evidence
that the absorber is a DLA system. 
Should the system have metallicity below solar, 
as generally observed in the QSO absorbers,  
the real $N$($\Ho$) would be even higher.

The column densities of the metals lend support to an origin
of the absorber in a galactic disk. 
In Figs. 3, 4 and 5 we compare the $\Fep$, $\Mnp$, $\Tip$, $\Cap$,
and $\Mgo$ column densities 
measured in the $z$=0.558 system with column densities observed
in Galactic interstellar clouds.  
Both the absolute and the relative values of the column densities of
the absorber are undistinguishable from those of
interstellar clouds within the disk of the Galaxy.
 This   agreement 
between several independent quantities is remarkable since the
measured column densities and their ratios are affected by a large
variety of factors, including chemical composition, 
presence or absence of dust, and physical state of the absorber.
 The observed general agreement suggests that
the system   originates in the disk 
of a galaxy lying along the line of sight to PKS 0118--272. 
This origin is consistent with  the absorber being
a Damped Lyman $\alpha$ system.

\begin{table}
\caption{$\Ho$ column density of the $z$=0.5580 system.}  
\begin{flushleft}
\begin{tabular}{cccccc}
\hline
  &  & \multicolumn{2}{c}{Base depletion } & 
      \multicolumn{2}{c}{Dense depletion } \\
Ion  & $N$($\Ho$)$_{\rm min}$  
& $\delta_{\rm X}$ & $N$($\Ho$)$_{\rm est}$ 
& $\delta_{\rm X}$ & $N$($\Ho$)$_{\rm est}$  \\
         & (cm$^{-2}$)  &  &  (cm$^{-2}$) &  &  (cm$^{-2}$)       \\
\hline 
$\Tip$   &   19.3   & --1.02 & 20.3 & --2.70 &  22.0 \\
$\Mnp$ &   19.4   & --0.84 & 20.2 & --1.28 &  20.7 \\
$\Fep$   &   19.0   & --1.27 & 20.2 & --2.03 &  21.0 \\
\hline
\end{tabular}
\end{flushleft}
\end{table}

\section{Search for an intervening galaxy in the field}

About 15 faint galaxies in the range of magnitude m$_R$ = 20.5 to 21.5,
are present within  30\asec   of the BL Lac object
(Falomo 1996). A close up of this field is shown in Fig. 2a. The
galaxy  closest to PKS 0118-27 lies at 8\asec (PA $\sim$
40$^{\circ}$), corresponding to a projected distance of $\simeq$ 70 kpc
($H_0$ = 50 km s$^{-1}$ and $q_0$ = 0) at $z=0.558$. 
This would imply a relatively large  size for
 a galactic disk with $N$($\Ho$) $\approx$ 10$^{20}$ atoms cm$^{-2}$. 
 This fact prompted
us to search for a closer companion galaxy by
subtracting a scaled Point Spread Function (PSF) from the image of the
target. As PSF template we used an unsaturated star at $\sim$
30\asec NE which is within our field of view and has brightness
similar to the target. 
After alignement of the PSF template with the
BL Lac image we   subtracted several scaled PSF changing the
normalization factor. The final value was set  to have minimum
residuals within a radius of 5 pixels. 
This PSF subtracted image reveals the presence of a companion galaxy   
1.6\asec\ NE of  0118-272 (Fig. 2b). 
This companion object is similar 
in magnitude to the galaxy at  8\asec\ NE
which has m$_R$ = 21.1.
If they are at the same redshift of the absorber 
their absolute magnitude is  M$_R \approx$ --22.3 and the 
  closest companion lies at only 14 kpc 
(projected distance) from the BL Lac source.
These   magnitudes and   distances  
  are in line with the results found by Le Brun et al.\,(1997) 
from their study of seven fields around quasars showing
DLA absorption at moderate redshift.
These authors  
find galaxies with   --23  mag $\leq$ M$_B$ $\leq$ --19  mag
and impact parameters $\approx$ 10 $h_{50}^{-1}$ kpc.  
The faint companion object at 1.6\asec (Fig. 2b) is the most likely
candidate  galaxy responsible for the absorption.
The   magnitude of the close companion
galaxy suggests that we are dealing with a normal galaxy.
A column density  $N$($\Ho$) $\simeq$ 2 $\times$ 10$^{20}$ cm$^{-2}$
is within the range expected for a normal galaxy 
at a galactocentric distance of  14  kpc.
The detection of a close galaxy with such properties
lends further support to   the notion that the
$z$=0.558 absorber is a DLA system.

\section{Properties of the $z_{\rm abs}$ = 0.558 absorber} 

\subsection{Elemental abundances and depletions}

To study the elemental abundances we consider elements that have
  their dominant ionization states in $\Ho$ regions, such as
$\Tip$, $\Mnp$ and $\Fep$. For reasons mentioned in Sect. 4.1
we do not consider $\Mgp$ and $\Cap$.
Owing to the lack of a direct $N$($\Ho$) determination we focus
our attention on relative abundances, i.e. element-to-element ratios.
In Table 5 we compare iron peak abundances  
in  DLA systems, including the absorber
towards PKS 0118--272.  The usual definition 
[X/Fe] = log $N$(X)/$N$(Fe) -- log (X/Fe)$_{\sun}$
is adopted. 
Deviations from solar ratios are generally found
in the absorbers. The interpretation of such relative abundances, 
however, is not
straitghforward since depletions on dust grains may distort the
true abundance ratios. 
In order to draw conclusions about abundances and depletions
 we need a set of fiducial cosmic 
abundances at large look-back times and, in addition,  
a realistic model for the pattern of elemental depletion. 
No conclusion could be
derived from the observed abundances should we allow
{\it a priori} an arbitrary chemical evolution and an
arbitrary elemental depletion pattern for the absorbers. 
Instead, by assuming a fiducial set of abundances and 
depletions,
 we can make quantitative predictions on the possible range
of observed metallicities.

\begin{table}
\caption{Iron peak relative abundances   in  
damped Ly\,$\alpha$ systems and in Galactic metal-poor stars.}  
\begin{flushleft}
\begin{tabular}{lccccc}
\hline
QSO  &  $z_{\rm abs}$ &  
$\left[ {\Fe \over \Hy} \right]_\obs$  &
$\left[ {\Mn \over \Fe} \right]_\obs$  &  
$\left[ {\Ti \over \Fe} \right]_\obs$  & Ref. \\
\hline 
0118--272 & 0.558 & ---  & +0.41 & +0.31 & 1 \\ 
2206--199 & 0.752 & --- &--0.10 & +0.27 & 2 \\
0454+039  & 0.860 & --1.0  &--0.36 & ---   & 3  \\
0014+813  & 1.112 &  --- &--0.10 & +0.42 & 4  \\
0450--132 & 1.174 & --1.5  &--0.22 & ---   & 3  \\
0449--134 & 1.267 & --1.5 &--0.28 & ---   & 3  \\
0935+417  & 1.373 &  --1.0 &--0.39 & +0.26 & 5  \\
0216+080  & 1.768 &  --1.0 &--0.07 & ---   & 3  \\
0528--250 & 2.141 &  --1.3 &--0.58 & ---   & 3  \\
\hline
\hline
   &  & 
$\left[ {\Fe \over \Hy} \right]_*$  & 
$\left[ {\Mn \over \Fe} \right]_*$  &  
$\left[ {\Ti \over \Fe} \right]_*$ & Ref. \\
\hline 
Stars with  &  &  --1.0 &  --0.35 & +0.22 & 6 \\
Stars with  &  &  --2.0 &  --0.42 & +0.30 & 6 \\
\hline
\end{tabular}

\smallskip

[X/Fe]$_\obs$: apparent metallicity not corrected for possible effects of dust
(see text for more details). 

\smallskip 

(1) Present work; (2) Prochaska \& Wolfe (1997);
(3) Lu et al. (1996); (4) Roth \& Songaila (1997; in preparation) ; 
(5) Meyer et al.\,(1995);
(6) Ryan et al. (1996). 
\end{flushleft}
\end{table}

The most natural reference for early galactic abundances comes 
from the study of 
metal-poor stars in the  Galactic halo (Ryan et al. 1996, and refs.
therein). 
In the last rows of Table 5 we give the typical Mn and Ti abundances
relative to Fe found  by Ryan et al. (1996) at metallicities [Fe/H] = --1
and --2. 
The [X/Fe]$_*$ values given in the table have been
estimated from the midmean vector defined by Ryan et al. (1996)
and have a scatter of 0.1 to 0.2 dex.

For the  depletion pattern, we assume that the dust has the same
general properties as Galactic interstellar dust and we adopt 
the representative $\delta_{\X,\ism}$ values given by Jenkins (1987). 
By analogy with the   
interstellar medium, 
we define the depletion of an element X in the QSO absorber as 
$
\delta_{\X,\dla} = \log\,N({\rm X})/N({\rm H}) - \log\,({\rm X/H})_{\dla} .
$
The intrinsic   element-to-element ratios in the DLA,
${\rm [X/Fe]}_\dla$, are related to the observed ratios,
${\rm [X/Fe]}_\obs$, by means of the relation 
$
{\rm [X/Fe]}_\dla = 
{\rm [X/Fe]}_\obs - (\delta_{\X,\dla}  - \delta_{\Fe,\dla} ) .
$
Since the quantities $\delta_{\X,\dla}$ are unknown 
we assume $\delta_{\X,\dla}$ = $\delta_{\X,\ism}$ and
we compute
$
{\rm [X/Fe]}_{\rm est} = 
{\rm [X/Fe]}_\obs - (\delta_{\X,\ism}  - \delta_{\Fe,\ism} ) ~,
$
which represents an estimate of the metallicity after correction
for dust effects.
In Table 6 we give the   [X/Fe]$_{\rm est}$ values in two cases:
(1)   base depletions and (2) dense depletions,
representative of interstellar lines of sight with a
low and a high amount of dust, respectively.

\begin{table}
\caption{Predicted relative abundances after correction for dust} 
\begin{flushleft}
\begin{tabular}{lcccccc}
\hline
& & \multicolumn{2}{c}{Base depletion$^1$} 
  & \multicolumn{2}{c}{Dense depletion$^2$} \\
QSO & $z_{\rm abs}$ & 
$\left[ {\Mn \over \Fe} \right]_{\rm est}$  & 
$\left[ {\Ti \over \Fe} \right]_{\rm est}$  & 
$\left[ {\Mn \over \Fe} \right]_{\rm est}$  &  
$\left[ {\Ti \over \Fe} \right]_{\rm est}$   \\
\hline  
0118--272 &  0.558 &   --0.02 & +0.06  &  --0.34   &  +0.98 \\
2206--199 &  0.752 &   --0.53 & +0.02  &  --0.85   &  +0.94 \\
0454+039  &  0.860 &   --0.79 & ---    &  --1.11   &   ---  \\
0014+813  &  1.112 &   --0.53 & +0.17  &  --0.85   &  +1.09 \\
0450--132 &  1.174 &   --0.65 & ---    &  --0.97   &   ---  \\
0449--134 &  1.267 &   --0.71 & ---    &  --1.03   &   ---  \\
0935+417  &  1.373 &   --0.82 & +0.01  &  --1.14   &  +0.93 \\
0216+080  &  1.768 &   --0.50 & ---    &  --0.82   &   ---  \\
0528--250 &  2.141 &   --1.01 & ---    &  --1.33   &   ---  \\
\hline
\end{tabular}

\smallskip

[X/Fe]$_{\rm est}$ = [X/Fe]$_\obs$ -- ($\delta_{\Mn,\ism}$~$-$~$\delta_{\Fe,\ism}$)
 
\smallskip 

$^1$ Dust with Galactic base depletion pattern: \\
($\delta_{\Mn,\ism}$~$-$~$\delta_{\Fe,\ism}$) = +0.43
and
($\delta_{\Ti,\ism}$~$-$~$\delta_{\Fe,\ism}$) = +0.25.

\smallskip

$^2$ Dust with Galactic dense depletion pattern:\\
($\delta_{\Mn,\ism}$~$-$~$\delta_{\Fe,\ism}$)= +0.75
and
($\delta_{\Ti,\ism}$~$-$~$\delta_{\Fe,\ism}$)= --0.67.

\end{flushleft}
\end{table}

The potential presence or absence of dust
and the possibility of a solar- or of a halo-type abundance 
pattern combine
to give four possible basic pictures:

\begin{itemize}

\item{\it No dust and solar-like abundances.}
In this case it should be
[X/Fe]$_\dla$~= [X/Fe]$_\obs$ = 0 for both Mn and Ti. 
This possibility is excluded from an inspection of Table 5. 
We are forced to invoke dust and/or chemical evolution effects. 
 
\item{\it Dust and  solar-like abundances.}
In this case the metallicities should be solar after
correction for dust effects, i.e. [X/Fe]$_{\rm est}$ $\simeq$ 0 
for both Mn and Ti. 
As can be seen in Table 6, this condition is satisfied only by
the $z=0.558$ absorber in PKS 0118--272 
for dust with  base depletion pattern.
None of the absorbers satisfies this condition if a dense
depletion pattern is assumed. 

\item{\it No dust and halo-like abundances.}
The metallicity is given directly
by [X/Fe]$_\obs$,  without correction for dust; 
if the chemical history of the absorbers follows  that
of our Galaxy, then should be
[X/Fe]$_\obs$   $\simeq$ [X/Fe]$_*$ in Table 5.
With the exception of the  
$z=0.558$ system, all the QSO absorbers   show   the typical
trends of metal-poor stars, i.e. [Mn/Fe] $<$ 0 and, when
available, [Ti/Fe] $>$ 0. 
In particular, the $z=1.373$ system
in QSO 0935+417 matches remarkably well the Galactic halo abundances
of both elements. 
 
\item{\it Dust and halo-like abundances.}
In this case the metallicities corrected for dust should match
those of metal-poor stars, i.e.[X/Fe]$_{\rm est}$ $\simeq$ [X/Fe]$_*$
in Table 6.
The $z=1.112$ absorber in QSO 0014+813 approximately matches
the condition for base elemental depletion. 
No absorbers  fulfil this
requirement for a dense depletion pattern.

\end{itemize}

In summary, in the sub-sample of  QSO absorbers considered
we have some cases consistent with dust-free, metal-poor gas
(for example the $z=1.373$ system in QSO 0935+417), 
one case consistent with metal-poor gas with dust
(the $z=1.112$ absorber in QSO 0014+813), and finally
the $z=0.558$ system under investigation which is the only
case indicating gas plus dust with overall
solar-like composition. 
The full set of observations is consistent
with a simple evolutionary scenario in which at early cosmic time
the gas is metal-poor and the dust not yet developed, at an intermediate
phase the dust starts to form and, finally, closer to the
present time, the absorber has attained an overall
(gas plus dust) solar composition. 
If this evolutionary
interpretation is correct, the QSO absorbers
are migrating in   Fig. 3  
from the region below the dashed line (i.e. the solar ratio), where dust 
is absent
and the gas is metal-poor, to the region above the dashed line,
where dust has already been formed and the metallicity is
solar-like. 
This scenario is consistent with the suggestion by Lu et al. (1996)
that high-redshift DLA systems ($z \geq 2$) are essentially dust-free and 
show a halo-like pattern of metallicities. 
However, this interpretation has problems in explaining the full set
of observed abundances in DLA systems. For instance, 
the measured [Zn/Fe] ratios do not follow the expected halo-like trend.
Moreover, the nitrogen
over-abundances and the [S/Zn] solar ratios 
observed in some DLA systems 
(Molaro et al. 1996, 1997)  are
contrary to those expected at low metallicities and suggest
that these systems may have a peculiar chemical history 
(Matteucci et al. 1997). 
In addition,  
there are indications that dust is present also at high redshifts,
albeit at a dust-to-gas
ratio about 10\% of the Galactic value
(Pettini et al. 1997; Fall \& Pei 1989). Finally,
deviations from the basic evolutionary scenario can  be expected if  
DLA systems include galaxies of different morphological type,
as indicated by the study of Le Brun et al. (1997).

\begin{figure}   
\psfig{figure=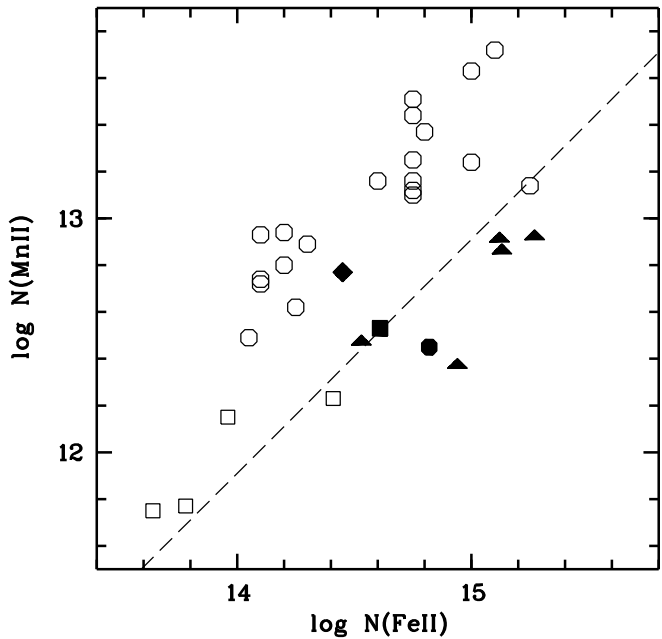,width=8.6cm}
\caption{
Filled symbols: $\Mnp$ and $\Fep$ column densities in DLA systems.
Diamond: absorber at $z$=0.558 towards PKS 0118-272.
Square: absorber at $z$=0.752 towards QSO 2206--199 (Prochaska \& Wolfe, 1997).
Triangles: DLA systems from Lu et al.\,(1996).
Octagon: absorber at $z$=1.373 in QSO 0935+417 (Meyer et al. 1995).
Empty symbols: $\Mnp$ and $\Fep$ column densities
in Galactic interstellar clouds. 
Empty octagons: interstellar clouds from the survey of Jenkins et al.\,(1986).
Empty squares: high velocity clouds towards HD 93521 
(Spitzer \& Fitzpatrick 1995).
Dashed line: solar Mn/Fe abundance ratio from Anders \& Grevesse (1989),
with iron taken from Hannaford et al. (1992).
}
\end{figure}

\subsection{Dense vs. base dust depletion}

This study suggests that the dust, when present,
has properties typical of low-density lines of sight, i.e. base depletion
rather than dense depletion pattern. This  
might be the result of a selection bias:  
if the gas with base depletion has a larger
cross section  than the gas with dense
depletion, then a sample of lines of sight randomly distributed through the
absorbers will preferentially intersect gas with 
low depletion.  
A larger cross section of gas with low depletion 
is observed in the Galaxy, where dense depletion is generally found
in lines of sight crossing clouds with
small geometrical filling factors.

\subsection{[Mn/Fe] and [Ti/Fe] ratios as discriminants between dust
and chemical evolution}

As pointed out by Lu et al. (1996),
the opposite behaviour of manganese and iron with respect to
differential depletion (\,[Mn/Fe] $>$ 0 in the ISM) and to
chemical evolution (\,[Mn/Fe] $<$ 0 in halo stars) makes the
[Mn/Fe] ratio a good diagnostic for disentangling the two effects. 
The $z=0.558$ absorber is the first example which shows
clear evidence of dust on the basis of its positive [Mn/Fe] ratio.
In principle,  [Ti/Fe] is also a good discriminant, since
titanium is usually more depleted than iron and so
[Ti/Fe] $<$ 0 in the ISM, while 
[Ti/Fe] $>$ 0 in halo stars. 
However, the slope of the interstellar correlation
of $\delta_{\Ti,\ism}$ with the average gas density along the
line of sight is much steeper than the slope of  the
$\delta_{\Fe,\ism}$ correlation (Jenkins 1987).  
Thus, in lines of sight with low density  
 titanium is less depleted than iron and
[Ti/Fe] $>$ 0, as in halo-like abundances.  
Therefore,
caution should be used in using the [Ti/Fe] ratio to unravel
dust depletion from metallicity evolution, 
especially  if base depletion is more commonly
detected in the absorbers, as discussed in the previous paragraph. 
The positive [Ti/Fe] ratio in the $z_{\rm abs}$ = 0.558 system
is perfectly consistent with the presence of dust
with no need to invoke a halo-like abundance pattern.

\begin{figure}   
\psfig{figure=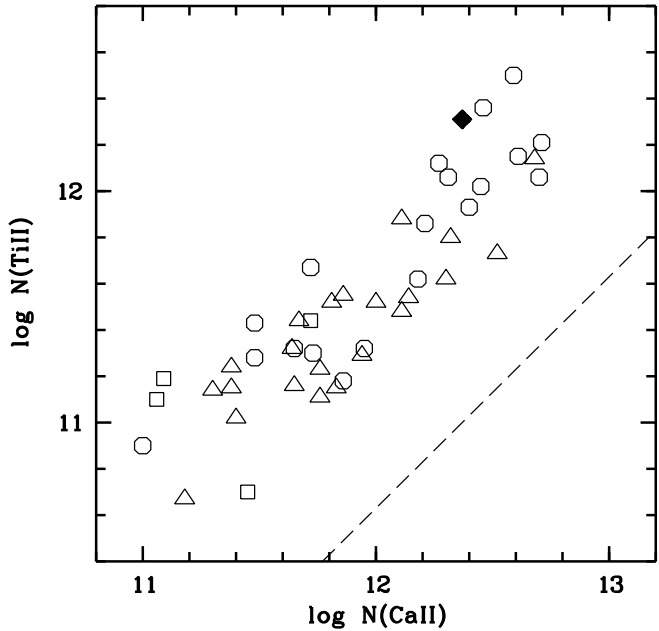,width=8.6cm}
\caption{
Comparison of the $\Cap$ and $\Tip$ column densities measured
in the absorption system at $z$=0.558 towards PKS 0118-272 (filled diamond)
and in Galactic interstellar clouds  (empty symbols).
Dashed lines: solar abundance Ti/Ca ratio  from
Anders \& Grevesse (1989).
Empty triangles: Galactic disk interstellar clouds from the 
compilation
of Crinklaw et al. (1994). 
Octagons: Galactic halo lines of sight from the survey of Albert et al. (1993). 
Squares: high velocity clouds towards HD 93521 
(Spitzer \& Fitzpatrick 1995). 
}
\end{figure}

\subsection{$N(\Tip )$ vs. $N(\Cap)$ correlation}

The $z_{\rm abs}$ = 0.558 system towards PKS 0118--272
is the only QSO absorber   with both $\Cap$ and $\Tip$
column density measurements. In Galactic interstellar clouds
there is a well known correlation between $N$($\Cap$) and $N$($\Tip$)
(Albert et al. 1993 and refs. therein). 
In Fig. 4 we compare $N$($\Cap$) and $N$($\Tip$)
in our system and in Galactic   clouds. 
Considering the high spread of calcium and titanium  
depletions and the possible presence of ionization effects
($\Capp$ can also exist in $\Ho$ regions), it is remarkable that
the point corresponding to PKS 0118-27  
matches the interstellar $\Tip$ - $\Cap$ correlation.
This result  reinforces the conclusion that the absorber
  originates in the interstellar gas of an intervening galaxy. 
Chemical evolution effects are not expected since   Ti and
Ca show similar trends in Galactic  halo stars, i.e.
[Ca/Fe]$_*$ $\simeq$ [Ti/Fe]$_*$ (Ryan et al. 1996).

\begin{figure}   
\psfig{figure=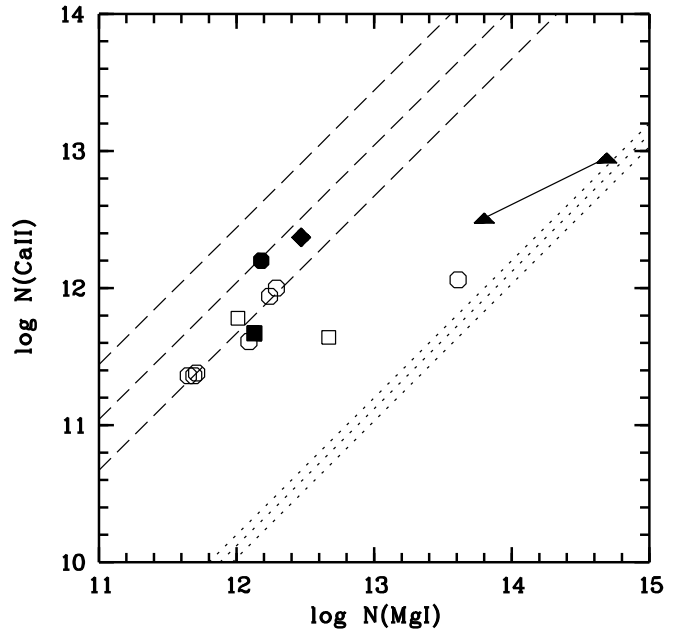,width=8.6cm}
\caption{ 
Filled symbols: $\Cap$ and $\Mgo$column densities in QSO absorbers.
Diamond:  
$z_{\rm abs}$ = 0.559 absorber towards PKS 0118-272.  
Triangles connected by a segment: measurements of the
$z_{\rm abs}$ = 0.692 absorber towards 3C 286 (Cohen et al. 1994);
Square:
$z_{\rm abs}$ = 0.5240 absorber in QSO 0235+164 (Lanzetta \& Bowen 1992;
velocity component with column density error $<$ 0.2 dex).  
Empty symbols: Galactic interstellar clouds. 
Octagons: Galactic components in the direction of SN1993J; 
$N$($\Cap$) from Vladilo et al. (1994); $N$($\Mgo$) from Bowen et al. (1994);
only components coincident within $\pm$ 3 km s$^{-1}$
in velocity are considered.  
Squares: interstellar components towards HD 149881 
(Spitzer \& Fitzpatrick, 1995). 
Broken lines: predicted  ${N(\Cap) / N(\Mgo)}$   ratios
as computed from Eq. (5) in the appendix. 
Dotted lines: models for cold gas (log\,$T_{\rm K}$ = 1.8,1.9,2.0 
from bottom to top) 
and high differential depletion  ${(\delta_{\Cap}-\delta_{\Mgp})}$ = --2.8 dex.
Dashed  lines: models 
for warm gas (log\,$T_{\rm K}$ = 3.8, 3.9, 4.0 from top to bottom)   
and low differential depletion ${(\delta_{\Cap}-\delta_{\Mgp})}$ = --1.7 dex. 
See Appendix for more details. 
}
\end{figure}

\subsection{$N(\Cap)/(\Mgo)$ ratio and physical state}

The ratio between $\Cap$ and $\Mgo$ column densities is quite
sensitive to the physical parameters of the gas. 
In the Appendix we derive an expression for the
 ${N(\Cap) / N(\Mgo)}$ ratio, 
which is valid for Galactic interstellar gas.  
The  ratio should be insensitive
to chemical evolution effects since [Ca/Fe]$_*$ $\simeq$ [Mg/Fe]$_*$
in metal-poor stars  (Ryan et al. 1996). 
In Fig. 5 we compare measured $\Cap$ and $\Mgo$column densities
with predicted values of the  ${N(\Cap) / N(\Mgo)}$ ratio
based on the equation derived in the Appendix.
Measurements include the few QSO absorbers with both species measured 
(filled symbols) and Galactic interstellar clouds observed  
at high  spectral resolution (FWHM $\simeq$ 5 km s$^{-1}$) both
in $\Cap$ and in $\Mgo$ (empty symbols).
The data points
corresponding to low redshift absorbers
are not separated from, but rather overlapping with
those corresponding to interstellar gas. 
This suggests qualitatively   that the physical conditions in
the QSO absorption systems are similar to those 
found in interstellar clouds. 
The measured $N(\Cap)/N(\Mgo)$ ratios span 
at least one order of magnitude, but there is some evidence
of grouping around some specific values. 
Three out of the 4 QSO absorbers (including our system),  
together with the Galactic 
high-velocity components   to SN1993J or HD 149881
have  $N(\Cap)/N(\Mgo)$ $\simeq$ --0.2. 
The absorber
at $z_{\rm abs}$ = 0.692 in  3C 286 and  
the  low-velocity SN1993J component  originating in
the Galactic disk  have instead  $N(\Cap)/N(\Mgo)$ $\simeq$ --1.4.
The predicted  $N(\Cap)$/$N(\Mgo)$ ratios are shown in Fig. 5
as dotted lines (cold gas with dense depletion pattern) and
dashed lines (warm gas with base depletion). 
The $N(\Cap)$/$N(\Mgo)$ ratio
is expected to be higher in warm gas with low depletion  than  
  in cold gas with high depletion. 
The models of  warm gas with low depletion naturally match  
the Galactic high velocity components  of SN1993J, while
  the models of cold gas with high depletion
match the disk component of SN1993J. 
If the same results can be applied to the QSO
absorbers, then the 3 absorbers with relatively 
high $N(\Cap)/N(\Mgo)$ ratio
are associated with warm gas and low depletion,
whereas  the  $z$ = 0.692 absorber in 3C 286  
is associated with cold gas and high depletion.  
This is consistent with the hypothesis that lines of sight
through highly depleted gas are less common than lines
of sight with low depletion (see 6.2).

\subsection{Kinematics and location of the absorber}

Multiple components with velocity spread up to a few
hundred km s$^{-1}$ are frequently found in DLA systems and
are interpreted as a signature of a rotating galactic disk.
The simple kinematical structure of the absorber (Sect. 3.1) 
does not exclude
that the line of sight is intersecting a galactic disk.
In fact, if the galaxy is seen face on, 
the profile is not be expected to be broadened by disk rotation.  
Also the absorber at $z$ = 1.3726  towards QSO 0935+417,
one of the other two DLA systems at moderate redshift studied at high 
resolution, shows a simple kinematical structure. Therefore,  
the lack of multiple components does not contradict 
our conclusion that the $z$=0.558 absorber  is a DLA system. 
 
The  detection of $\Cap$ and $\Tip$ gives some indication on the 
location of the absorbing gas in the intervening galaxy. 
In our Galaxy 
the scale heights of $\Cap$  and  $\Tip$
above the Galactic plane  
are  $\approx$ 1  kpc and $\leq$ 1.5 kpc  respectively
(Albert et al. 1993; Lipman \& Pettini 1995).
This is consistent with an origin in the disk  
of the galaxy intercepted by the
line of sight.

The observations of PKS 0118--272 are consistent with 
the presence of a luminous host (M$_R$ = --23.5) 
of the BL Lac. 
We do not believe, however, that the absorber is associated with the
host  galaxy. 
Host galaxies of BL Lac objects are generally elliptical or 
bulge-dominated systems of
absolute magnitude comparable with that of giant ellipticals
(Falomo 1996 and refs. therein). 
Even though several interstellar tracers 
have been detected in elliptical galaxies, 
there is no evidence that {\it neutral hydrogen} is a common
interstellar constituent of ellipticals  
(Roberts et al. 1991). This fact argues against an origin
of the absorber in the host galaxy. 
In addition, if the absorbing gas were located close to the source,
its ionization degree would   probably be affected by the 
extremely high luminosity of the host ($M_{\rm R}$ $\simeq$ $-23.5$ mag; 
Falomo 1996) and of the BL Lac active nucleus itself. 
On the contrary,  
the general properties of the system are indistinguishable from
those of interstellar clouds embedded
in the radiation field of our Galaxy, as mentioned in the
previous paragraphs.

\section{Summary and future work}

Studies of Damped Ly $\alpha$ systems at moderate redshift  ($z \leq 1.5$)
are quite rare, but important for linking the properties 
of high redshift DLA systems with those of present-day galaxies.
We expect a gradual variation of the properties with  
redshift (i.e. cosmic time) as a consequence of galactic
evolution. 
From the present study we find that  the  $z$ = 0.558 absorber 
in the direction of PKS 0118--272 
is most likely a Damped Ly\,$\alpha$ system at low redshift since
our indirect estimates of the hydrogen column density yield
a conservative lower limit
$N$($\Ho$)  $>$ 2.5 $\times$  10$^{19}$ cm$^{-2}$
and, more likely,
$N$($\Ho$)  $\simeq$ 2 $\times$ 10$^{20}$ cm$^{-2}$.
The system   most likely originates in a galactic
disk in front of PKS 0118-272, a conclusion supported  
by the following results:

\begin{itemize}
 
\item
We detect a galaxy at only 1.6\asec\ from the BL Lac object
by applying a PSF subtraction technique to our imaging data. 
If this close companion is at the
redshift of the system, its projected distance is 14 $h_{50}^{-1}$ kpc, 
consistent with typical impact parameters of
companion galaxies associated with DLA systems
(Le Brun et al. 1997).  
 
\item
The  $\Mnp$, and $\Fep$ column densities
and their ratios are remarkably similar to corresponding values  
measured in Galactic interstellar clouds.
The $\Tip$ and $\Cap$ column densities obey
the well known interstellar correlation between these
two quantities, the first evidence 
of this kind for a QSO absorber. 
The $\Cap$ and $\Mgo$ column densities and their ratio,
which is sensitive to physical conditions in the absorbing gas,
are in line with measurements 
and theoretical predictions for
  interstellar clouds. 
These results, together with the relatively low scale height of
$\Cap$ and $\Tip$ in our Galaxy, suggest that
the absorber originates in a galactic disk. 

\end{itemize}
 
We compared the [Mn/Fe] and [Ti/Fe] ratios measured in our system
with similar measurements in QSO absorbers, in halo stars
and in Galactic insterstellar gas. Since depletion on dust grains
acts as a disturbance for measuring the intrinsic metallicity, 
we corrected the ratios measured in the DLA systems from the effects of dust
by using  values of Galactic differential depletion   
in two representative cases:
 base depletion, typical of lines of sight with low
density, and  dense depletion, typical of high density
interstellar clouds. The outcome of this comparison
can be summarized as follows:

\begin{itemize}

\item
For the first time in a QSO absorber we find [Mn/Fe] $>$ 0, 
a clear signature of dust.
Previous studies of DLA
absorbers have found  [Mn/Fe] $<$ 0, 
a result which has been interpreted as   signature
of a halo-like metallicity pattern. 

\item
We find a positive [Ti/Fe] ratio, similar to the   measurements
of this ratio in DLA absorbers. However, we show
 that
the [Ti/Fe] ratio is not well suited for discriminating between
dust and chemical evolution.

\item
Lines of sight with a dense depletion pattern appear to be
uncommon among DLA absorbers.

\end{itemize}
  
With the limitations due to the small size of the sample,
the data seem to be consistent with a simple evolutionary scenario
in which at early epochs 
the absorbers are characterized by gas with low dust content, while closer to
the present time  
gas and dust coexist like in the Galaxy.  
The $z=0.558$ absorber towards PKS 0118--272 could be
the first DLA system for which we start to see the transition
to present-day abundance and dust properties. 
A more realistic correction of DLA abundances from dust effects,
able to take into account variations of metallicities and dust-to-gas ratios
among the absorbers, is required to derive firm conclusions 
on the chemical evolution of the intervening galaxies (Vladilo 1997). 
Only by accumulating abundance measurements for a large
number of DLA systems at high and low redshifts it will
be  possible to disentangle the different causes that combine to
produce the observed properties of these systems, which include
chemical evolution, dust  evolution, 
spatial abundance gradients, filling factors of interstellar phases, 
and differences in  morphological types of the DLA galaxies.

Ultraviolet spectroscopy of PKS 0118--272 is required to perform
a direct measurement of the $\Ho$ column density from  
  the Ly  $\alpha$ absorption profile; this will allow
the absolute metallicity of the system to be determined. 
High spatial resolution spectroscopy of the 
the newly discovered
companion at 1.6\asec\ from PKS 0118--272 is needed to
measure its redshift and 
confirm its association with the absorber.
Finally, high quality imaging from space  
is required to establish the morphological type of the  
companion.  This would give
the rare opportunity of linking  the galactic morphological type to 
the spectroscopic properties of a DLA system.

\acknowledgements{We thank J. Bergeron,  
D. Meyer, and K. Roth for providing results in advance of publication.}
 
\bigskip

\appendix{\bf Appendix: The ${ N(\Cap) / N(\Mgo) }$ interstellar
ratio. }
The  interstellar depletion of an element X 
in a diffuse medium with overall solar metallicity
is usually defined as: 
\begin{equation}
\delta_{\rm X} = \log { N({\rm X}) \over N(\h) } - 
\log ( { \X \over \h } )_{\sun}
\end{equation}  
where $N(\X)$ and $N(\h)$ are the   
column densities of X and of hydrogen in the gas phase, and
$(\X/\h)_{\sun}$ is the solar abundance. It is assumed that
the fraction of the element not present in gas phase,
which gives rise to a logarithmic depletion $\delta_{\rm X} < 0 $, is in dust. 
From (1) one has:
\begin{equation}
{ N({\rm Ca}) \over N({\rm Mg}) } =
({ {\rm Ca} \over {\rm Mg} })_{\sun}     
10^{( \delta_{\rm Ca} - \delta_{\rm Mg} )}    
\end{equation}
where $N({\rm Ca})$ and $N({\rm Mg})$ represent the sum over all
ionization states of Ca and Mg atoms in gas phase. 
In interstellar $\Ho$ regions, the fraction of neutral stages of the metals
are generally negligible since they are  ionized by  
photons with h$\nu$ $<$ 13.6 eV and one can write
$
N({\rm Ca}) = N(\Cap) + N(\Capp)  
$
and
$
N({\rm Mg}) = N(\Mgp)  .
$
From these relations we get
$
\delta_{\rm Ca} = \delta_{\Cap} + \log~\{~1~+~[~N(\Capp)~/~N(\Cap)~]~\}
$
and
$
\delta_{\rm Mg} = \delta_{\Mgp}  .
$
Even if $\delta_{\Cap}$ does not represent the true depletion
of calcium from gas to dust, it is a useful quantity for
comparison with literature data.  From the above relations we obtain
\begin{equation}
{ N(\Cap) \over N(\Mgp) } = 
({ {\rm Ca} \over {\rm Mg} })_{\sun} 
  10^{( \delta_{\Cap} - \delta_{\Mgp} )} ~.
\end{equation}
 
The ionization equilibrium equation for magnesium in $\Ho$ regions is:
\begin{equation}
{ N(\Mgp) \over N(\Mgo) } =  
{ \Gamma_{\Mgo} ~+~ C_{\Mgo}(T)   n_e \over
\alpha_{\Mgp}(T)   n_e }
\end{equation}
where $\Gamma$ is the photoionization rate, $C(T)$ the collision
ionization rate, $n_e$ the electron density, and $\alpha(T)$
the total recombination coefficient (radiative plus dielectronic). 
By multiplying the last two equations we finally obtain
\begin{equation}
{ N(\Cap) \over N(\Mgo) } = 
({ {\rm Ca} \over {\rm Mg} })_{\sun}  
10^{(\delta_{\Cap}-\delta_{\Mgp})}      
{ \Gamma_{\Mgo} ~+~ C_{\Mgo}(T)   n_e \over
\alpha_{\Mgp}(T)   n_e }
\end{equation}
This expression allows us to predict the value of the ratio
${ N(\Cap) / N(\Mgo) }$    for a range
of  physical parameters and depletions representative
of interstellar conditions. 
We adopted the  magnesium photoionization rate $\Gamma_{\Mgo} =
3.6 \times 10^{-11}$ s$^{-1}$ from Keenan (1984),
the recombination coefficient $C_{\Mgo}(T)$  from
Shull \& Van Steenberg (1982), the direct ionization rate $\alpha_{\Mgp}(T)$
from Arnaud \& Rothenflug (1985), and an electron density 
$n_e$ = 0.03 cm$^{-3}$. 
The ${ N(\Cap) / N(\Mgo) }$ ratio generally
scales directly as the inverse of the electron density because
 collisional ionization
is negligible compared to  photoionization  
for the range of temperatures considered 
($T \leq 10^4$ K).   
Representative values of $\delta_{\Cap}$ are computed from the compilation of
Crinklaw et al. (1994) by using the definition of  base
and dense depletions given by Jenkins et al. (1987).
We find $\delta_{\Cap}$ = --2.13 (base depletion) and 
$\delta_{\Cap}$ = --3.79
(dense depletion). 
Corresponding values of $\delta_{\Mgp}$  
are taken directly from Jenkins et al. (1987):
$\delta_{\Mgp}$ = --0.16 for base depletions and  $\delta_{\Mgp}$ = --0.72
for dense depletions. These $\delta_{\Mgp}$ values are derived from a 
survey of the $\Mgp$ 1240\AA\ doublet (Jenkins et al. 1986)
and should be increased by --0.67 dex if the
revised oscillator strength  of the 1240\AA\
doublet is adopted (Sofia et al. 1994). 
For base depletion we have therefore 
$(\delta_{\Cap}-\delta_{\Mgp})$ = --2.0 (or --1.3 with the
revised oscillator strength), while for dense  depletion
we have $(\delta_{\Cap}-\delta_{\Mgp})$ = --3.1 (or --2.4).

\end{document}